# A PMU-based Multivariate Model for Classifying Power System Events

Rui Ma, *Student member, IEEE*, Sagnik Basumallik, *Student member, IEEE* and Sara Eftekharnejad, *Senior Member, IEEE*

*Abstract*--Real-time transient event identification is essential for power system situational awareness and protection. The increased penetration of Phasor Measurement Units (PMUs) enhance power system visualization and real time monitoring and control. However, a malicious false data injection attack on PMUs can provide wrong data that might prompt the operator to take incorrect actions which can eventually jeopardize system reliability. In this paper, a multivariate method based on text mining is applied to detect false data and identify transient events by analyzing the attributes of each individual PMU time series and their relationship. It is shown that the proposed approach is efficient in detecting false data and identifying each transient event regardless of the system topology and loading condition as well as the coverage rate and placement of PMUs. The proposed method is tested on IEEE 30-bus system and the classification results are provided.

*Index Terms*--Bag of Pattern (BOP), classification, Phasor Measurement Units (PMUs), Symbolic Aggregation approXimation (SAX)

## I. Introduction

Near real-time transient event detection and identification is essential in power system protection and control due to increased interconnectivity [1]. PMUs deployed in power systems enable real-time visualization of disturbance events as they generate synchronized voltage phasors, current phasors, and frequency measurements at a refresh rate of 30 to 120 messages per second that are stamped with a reference clock from Global Positioning System (GPS). This provides PMUs with the ability to capture power system transient events with a high accuracy.

Researchers have investigated using PMU data to detect and identify transient events such as faults, line tripping and generation loss [2]–[8]. Traditional signal processing techniques such as wavelet analysis, short time Fourier transform (STFT) and Hilbert analysis are utilized to analyze the oscillation mode of PMU signals under different transient events [2]–[4]. However, the signal processing methods require an appropriate sampling rate and a proper length of sample window to guarantee the event detection performance [9]. Authors in [5] and [6] utilize a multidimensional ellipsoidal method to enclose the PMU data into an ellipsoid so that features such as volume and orientation can be used to identify power system disturbances. A drawback of this method is that it requires significant time for classification [7]. Instead of using all the PMU data accumulated in phasor data concentrator (PDC), [1] and [8] utilize a single PMU's data with the largest variation to identify transient events quickly and precisely. The time series subsequences called shapelets [10], that can maximally represent a PMU data, are utilized to recognize disturbance pattern and enhance the accuracy of classification. Although the classification accuracy of the aforementioned methods is promising for detecting transient events, there is still room for improvement.

Before utilizing PMU measurements to classify transient events, the correctness of PMU data has to be verified. Authors in [11]–[14] have shown that a malicious cyber-attack on PMUs can remain undetected. Authors in [15] and [16] show that attackers can mimic a transient event to inject false data into a set of PMU streams. Hence, it is essential to distinguish false data from real transient events. The availability of PMU-generated time series data creates an opportunity to deploy classification techniques to identify this false data. Time-series classification techniques have been developed for pattern recognition and similarity search. The majority of classification techniques can be divided into two categories: shape-based methods and structural-based methods [17]. Shape-based methods compare the distance between two time series data points. Euclidean Distance and Dynamic Time Warping (DTW) are the two common shape-based techniques used to quantify similarity [18]. These methods perform well on short time series while the results on long-term time series are poor or require heavy computation time [19]. On the contrary, structural-based methods use global features such as sparsity or entropy to compare long-term time series' similarity instead of analyzing each point of time-series data. One of the widely used structural-based methods, Bag-of-Pattern (BOP), uses the sparsity of each time series to measure similarity [20]. Nevertheless, the aforementioned shape-based and structural-based classification methods focus on a univariate classification problem. Biswal et al. in [1] explain that the oscillation patterns of a power system disturbance recorded by different PMUs are similar, whereas the strength is different as it varies with the distance between the location of a disturbance and the PMUs. As a result, the transient event classification can be formed as a multivariate time series classification problem.

In this paper, a multivariate PMU time series classification mechanism is proposed to classify transient events and detect false data that can bypass state estimation. For each event, all time series data received from PMUs are pruned and represented as symbolic words. This is achieved by the Symbolic Aggregation approXimation (SAX) technique, where a time series can be divided into several equal-sized subsets and converted to a set of words based on the mean value of each subset. By counting the occurrence of a word appeared in the PMU

This work is supported by the National Science Foundation (NSF) Grant No. 1600058.

time series data, a multivariate BOP representation can be created to further extract the disturbance patterns shared by each PMU time series. Inspired by text mining, a modified document classification method, i.e. term frequency and document frequency (TF-DF), is developed to discover the oscillation patterns of PMU time series in each of the six types of transient events, i.e. false data injection, faults, line tripping, load change, generation loss, and shunt switching. Classification methods are used to classify these events based on features extracted from these patterns.

This paper is organized as follows: Section II describes the SAX technique used in this paper. The proposed multivariate Bag-of-Pattern method is introduced in Section III and modified document classification TF-DF model is discussed in Section IV. Section V presents case studies where the effectiveness of the introduced approach on classifying various types of transient events is investigated.

## II. Symbolic Aggregation approXimation (SAX)

Processing large volumes of PMU data is time consuming, which is not desired for near real-time event identification. Authors in [21] point out that 100 PMUs can generate up to 50 GB of data per day. This volume of data poses major challenges to hardware and software. Therefore, it is necessary to reduce the size of PMU data before identifying transient events. The SAX technique enables reducing dimensions of time series data without losing statistical features [20]. This technique can convert time series data into symbolic words so that text mining algorithms and string analysis methods can be used to efficiently analyze time series.

Given a PMU time series data $T$ of length $n$ and a sliding window of length $\omega$ ($\omega \leq n$), SAX converts $T$ into ($n - \omega + 1$) number of overlapping subsequences by continuously moving the window a unit distance ahead from the initial point of $T$. For each subsequence data $p$, SAX normalizes $p$ such that its mean is zero and its standard deviation is one. Then, the normalized subsequence $p$ will be divided into $\gamma$ number of equal-sized segments where the mean value of each segment is recorded. The number of segments $\gamma$ is defined as the size of a word. The subsequence $p$ can be represented by the recorded mean values. The process of aggregating each mean value to represent a subsequence $p$, known as Piecewise Approximate Aggregations (PAA), reduces the dimension of the data [22]. After forming PAA representation, each segment is converted into a letter by setting the alphabet size $\alpha$ and comparing the segment's mean value with a set of breakpoints that are presented in Table I [23]. Thereby, a symbolic word is constructed to represent the subsequence $p$. After transforming each subsequence into a symbolic word, the time series $T$ is converted into ($n - \omega + 1$) number of words.

Three parameters are required to convert time series into symbols: the alphabet size ($\alpha$), the size of a word produced by SAX ($\gamma$), and the length of the sliding window ($\omega$). The alphabet size is the number of alphabet letters that are used to represent time series. The word size is the number of symbols present in the output from the sliding window. The length of the sliding window is the number of points of time series that are to be con-verted. After defining these three parameters and following the breakpoints presented in Table I, a set of words are created as described in the aforementioned steps. For example, the time series of length 110 in Fig. 1 is converted to a symbolic word *bcbabddcaa* with SAX technique. The breakpoints are -0.67, 0 and 0.67 where alphabet size $\alpha$ is 4.

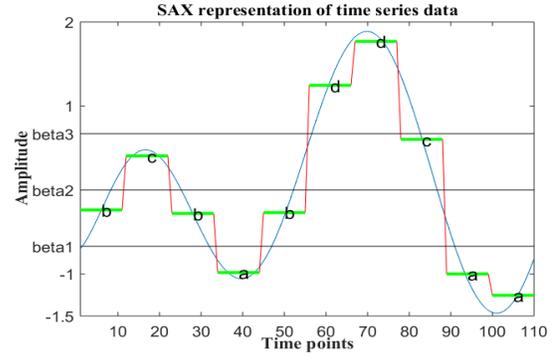

Fig. 1. Symbolic Aggregation approXimation ($\alpha$ = 4, $\gamma$ = 10, $\omega$ = 110, length of data = 110)

TABLE I
A LOOKUP TABLE THAT CONTAINS THE BREAKPOINTS THAT DIVIDE A GAUSSIAN DISTRIBUTION INTO 3 TO 5 EQUAL PROBABLE REGIONS

|  |  | Alphabet Size ($\alpha$) | | |
| --- | --- | --- | --- | --- |
|  |  | 3 | 4 | 5 |
| Break-point | $\beta_1$ | -0.43 | -0.67 | -0.84 |
|  | $\beta_2$ | 0.43 | 0 | -0.25 |
|  | $\beta_3$ |  | 0.67 | 0.25 |
|  | $\beta_4$ |  |  | 0.84 |

## III. Bag of Pattern

The SAX technique efficiently reduces the size of PMU time series and converts the data into words. In this paper, we convert transient event data received from PMUs into a set of words, that in turn are used by text mining techniques, such as Bag-of-Words (BOW) [24] for pattern extraction. The BOW technique, which is commonly used to classify univariate time series data, is an efficient way to capture the oscillation patterns of PMU time series data under different transient events. Since the problem in power system event identification is multivariate, a modified Bag-of-Pattern model, which is enabled by BOW, is utilized to convert a set of PMU time series data into a feature matrix such that the attributes of each PMU time series and the relationship between them can be identified.

### A. Bag-of-Words

BOW, which is inspired by information retrieval and text mining, has been applied to univariate time series classification in many prior works [17], [23]–[26]. Given the size of the alphabet $\alpha$, the size of the word $\gamma$ and the length of sliding window $\omega$, one can convert the univariate PMU time series $T$ of length $n$ into ($n - \omega + 1$) number of SAX words and then construct a frequency column vector of length $\alpha^\gamma$. Each element of this vector corresponds to a SAX word and the value of that element represents the frequency of the word occurring in the PMU time series. The frequency vector is the BOW representation of a univariate PMU time series $T$. One advantage of BOW is that the order of the pattern does not have to be considered.

TABLE II
BOW REPRESENTATION OF TIME SERIES SHOWN IN FIG. 1 (LEGTH OF DATA = 110, $\alpha = 3, \gamma = 2, \omega = 60$)

| SAX dictionary | Frequency |
|---|---|
| aa | 0 |
| ab | 0 |
| ac | 23 |
| ba | 0 |
| bb | 20 |
| bc | 0 |
| ca | 8 |
| cb | 0 |
| cc | 0 |

Given the time series in Fig. 1, one can convert the time series data into 51 subsequences by setting the $\omega$ to 60. Thus, 51 SAX words can be obtained. By counting the number of times each word occurs (e.g. the word *ac* occurs 23 times), the BOW vector, which is shown in Table II, is constructed. The alphabet size is equal to 3 and the size of the word is equal to 2, therefore, the size of the column vector is $3^2 = 9$.

*B. Multivariate Bag-of-Pattern*

When a transient event occurs in power systems, all PMU measurements will have a high correlation. On the contrary, a falsa data injection that results in a *fake transient* can only affect several PMUs where the measurements are less correlated. In order to successfully detect false data and classify transient events, a multivariate classification approach is developed to identify the attribute of each individual PMU measurement and the correlation between them. Authors in [23] propose a representation of Bag-of-Patterns (BOP) to analyze the similarity between each time series data. However, this only works for univariate time series classification while the classification problem in this paper is multivariate. A multivariate BOP representation is defined in [27] to capture the relationship of multivariate time series at a cost of losing the information of each individual time series data. The application of BOP has not been used in power system transient event identification to the best of our knowledge. We apply this BOP for power system transient events identification and extend the BOP to a multivariate level such that the information of each PMU time series data and the correlation between them are retained. A multivariate BOP matrix, which is the combination of several BOWs as illustrated in Table III, can be created to represent a multivariate PMU time series. Each row of the BOP matrix denotes a SAX word and each column corresponds to a PMU time series. The element of the BOP matrix represents the number of times a SAX word occurs in the corresponding PMU data. Intuitively, the oscillation patterns of the transient event signal hidden in each PMU time series are recorded and can be easily extracted.

IV. TERM FREQUENCY AND DOCUMENT FREQUENCY MODEL

The modified BOP representation enables transferring multivariate PMU time series data into a frequency matrix using SAX. However, the transient event pattern is not evident through the BOP matrix. In order to recognize the oscillation patterns of PMU time series hidden in the BOP matrix, a modified term frequency and document frequency (TF-DF) model is adopted in this paper. Traditional term frequency and inverse document frequency (TF-IDF) models introduced in [17] are deployed to classify univariate time series in a way that the words unique to a time series are highlighted. For a SAX word *t*, TF quantifies the number of times *t* occurred in a PMU time series while IDF represents the inverse fraction of the number of PMUs containing the word *t*. Contrary to the TF-IDF model, the TF-DF model aims to recognize the oscillation patterns of transient event signals that most PMU measurements record as these measurements are highly correlated.

TABLE III
BAG-OF-PATTERN REPRESENTATION OF MULTIVARIATE TIME SERIES ($\alpha = 3, \gamma = 2$)

| | | Multivariate PMU time series | | | | |
|---|---|---|---|---|---|---|
| | | PMU 1 | PMU 2 | PMU 3 | … | PMU $n$ |
| SAX dictionary | aa | 0 | 0 | 0 | … | 1 |
| | ab | 20 | 18 | 16 | … | 17 |
| | ac | 15 | 16 | 17 | … | 18 |
| | … | … | … | … | … | … |
| | cc | 0 | 0 | 2 | … | 0 |

Term frequency and inverse document frequency are scaled as follows [17]:

$$tf_{t,d} = \begin{cases} 1 + \log(f_{t,d}) & \text{if } f_{t,d} > 0 \\ 0 & \text{otherwiese} \end{cases} \quad (1)$$

$$idf_{t,D} = \log \frac{N}{|\{d \in D : t \in d\}|} = \log \frac{N}{df_t} \quad (2)$$

where $t$ is a SAX word extracted from a PMU time series, $d$ is a PMU time series, $f_{t,d}$ is the number of times $t$ occurs within the PMU time series $d$. $D$ represents a set of PMU time series data. $N$ is the total number of PMUs used to detect transient events, $df_t$ is the number of PMU time series that containing the SAX word $t$.

In a TF-IDF model, inverse document frequency (IDF) captures rare words as those are more informative than frequent words. However, the multivariate PMU time series classification problem tries to identify the word observed in most of the PMU measurements. Hence, a scaled document frequency (DF), which counts the number of PMU time series data containing word $t$, is defined as follows:

$$df_{t,D} = e^{\frac{|\{d \in D : t \in d\}|}{N} - 1} = e^{\frac{df_t}{N} - 1} \quad (3)$$

TF-DF is the product of two factors: term frequency and document frequency:

$$tf \times df = (1 + \log(f_{t,d})) \times e^{\frac{df_t}{N} - 1} \quad (4)$$

The TF-DF model efficiently identifies oscillation patterns of a multivariate PMU time series by assigning higher weight to the words that can represent most PMU time series. After transferring each PMU time series into a set of SAX words, TF counts the number of times a SAX word appeared in the corresponding PMU time series. DF counts the number of PMU time series that contain the SAX word. It converts the frequency matrix obtained by BOP into a weighted matrix such that the word that represents most of the PMUs within a multivariate PMU time series are assigned a higher weight while a lower weight is assigned to a word unique to a PMU time series data. Once the TF-DF weight matrix is constructed, features can be extracted to facilitate the classification. Given a BOP matrix in Fig. 2, the TF-DF weight matrix can be constructed based on (4). For example, in Fig. 2 the words *ba* and *bb* have higher weights in TF-DF matrix since the words *ba* and *bb* not only appear more often than the other words in each PMU time series but also appear in both PMU time series data in the BOP matrix.

Fig. 2 Converting BOP frequency matrix into TF-DF weight matrix

TABLE IV
NUMBER OF STUDIES EVENTS

| Event Type | Number of Events |
|---|---|
| Fault | 935 |
| Generation Loss | 115 |
| Line Tripping | 163 |
| Load change | 420 |
| Shunt Switching | 120 |
| False data injection | 600 |
| **Total** | **2353** |

## V. CASE STUDIES

The proposed modified BOP multivariate classification methodology has been studied on the IEEE 30-bus system and the method's applicability on larger systems is investigated. The accuracy of transient event classification is obtained using Support Vector Machine (SVM) and Ensemble classifiers [28], [29]. Different locations and coverage rates of PMUs are compared to analyze the effectiveness of the proposed approach. In addition, classification is investigated on noisy data to evaluate robustness of the developed method.

### A. PMU placement and measurements

PMU measurements outperform Supervisory Control and Data Acquisition (SCADA) measurements in terms of data quality and sampling frequency [8]. Although the penetration of PMUs in power systems has increased in recent years, the current deployment status of PMUs in most power systems is not having PMUs in each single location. Therefore, to study a practical system where part of the system is observed by PMUs, only 7 buses (buses 2, 4, 6, 10, 12, 15 and 27) are placed with PMUs in IEEE 30-bus system to provide 7 bus voltages and 28 branch currents at a rate of 30 samples per second. In our application, the time series of voltage and current measurements for the duration of 1.5 s (0.5 s of pre-event and 1 s of post-event data) are used to classify transient events and false data.

### B. Event creation

False data and five types of transient events, namely, faults, large generation loss, line tripping, large load change, and shunt switching are investigated in four different scenarios:

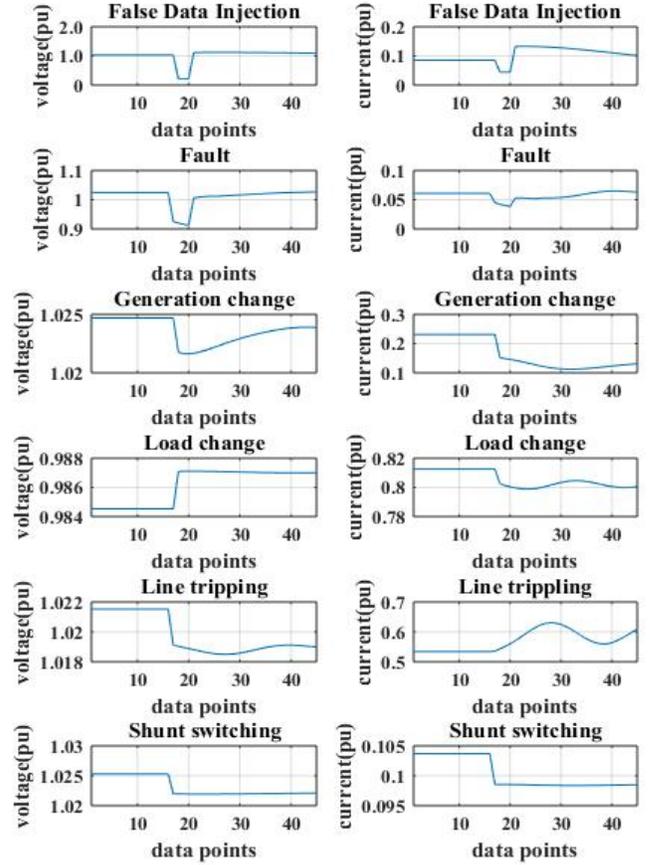

Fig. 3 Voltage and current waveforms for various event types

1) System with peak loading condition;
2) System with light loading condition;
3) System with light loading condition and one generator out of service; and
4) System with light loading condition and one line out of service.

A total of 2,353 events are generated as described in Table IV. An example of voltage and current waveforms for the six types of events are given in Fig. 3. The length of each waveform is 1.5 s, i.e. 45 points, to match PMU sampling rate of 30 frames per second. For each event, 7 PMU bus voltage and 28 PMU branch current magnitudes are used to construct multivariate BOP voltage and current matrixes that are further processed to convert to weighted matrixes using TF-DF algorithm.

### C. False data creation

Seamless power system operation is highly dependent on accurate state estimation. However, the accuracy of state estimation results is threatened by cyber-attacks. Authors in [11]–[13] indicate that attackers can bypass the existing bad data detection systems once they have the knowledge of the power system topology. In another work [14], it is shown that false data can be successfully injected into a hybrid state estimator where PMUs and conventional meters are used for state estimation. However, the mechanism of changing correct measurements to wrong measurements during an attack on a state estimator is not clearly defined. Although authors in [15] and [16] introduce a mirroring data spoof strategy to inject false data into PMU streams, the false data would be recognized as bad data in state estimation. This paper aims to detect the false data that can bypass bad data detection process in a hybrid state estimator, where both SCADA and PMU measurements are used.

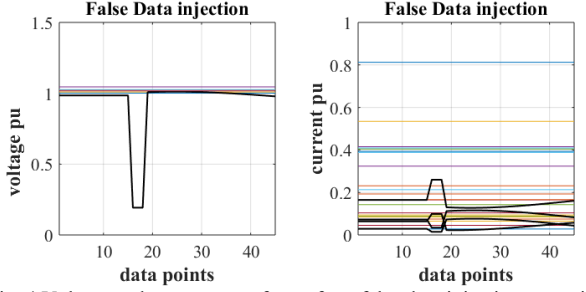

Fig. 4 Voltage and current waveforms for a false data injection scenario

The reporting rate of PMUs, which is 30 messages per second, is much faster than SCADA meters. Since state estimation runs every 30 seconds, it is necessary to use the average PMU measurements during this period to help estimate system states with the existing conventional measurements. PMU data buffer, proposed in [30], is applied to integrate PMU measurements into a hybrid state estimator. With this buffering model, attackers can compromise the 30 seconds data in a way that the waveforms of the recorded voltages and currents resemble a three-phase branch fault while the average values are changed. In our application, a damping equation is used to help represent a three-phase fault as:

$$V = e^{-at}(k \cdot \cos(bt+c) + k \cdot \sin(bt+c)) \quad (5)$$

where parameters $a$, $b$, $c$, and $k$ control the damping oscillation and magnitude.

The nature of a false data injection attack on PMUs is to change several PMU measurements to a specific value such that the state estimation converges to a wrong solution, while the remaining PMU measurements stay intact. Figure 4 illustrates an example of a false data injection where 7 bus voltages and 28 current magnitudes are plotted. The length of the voltage and current waveforms is 1.5s, which is 45 data points. Only a few voltage and current measurements are assumed to be falsified. This is true in a real power system due to the difficulty of compromising all PMU measurements.

### D. Feature extraction

The feature extraction process of a multivariate PMU time series data, which is plotted in Fig. 5, starts with applying the SAX algorithm to the labeled multivariate PMU current and voltage time series after defining the parameters $\alpha, \gamma$, and $\omega$. Taking a labeled multivariate PMU voltage time series as an example. Each PMU voltage time series is converted to a set of symbolic words by applying SAX method. Next, a BOP matrix $M_{BOP}$ is constructed where the number of the columns equals to the number of PMU voltage time series and the size of each column equals to $\alpha^\gamma$. The element $M_{BOP_{i,j}}$ denotes the number of times a SAX word in row $i$ occurs in PMU voltage time series $j$. Thus, the oscillation patterns of the PMU voltage time series are converted to frequencies and stored in a BOP matrix $M_{BOP}$.

After forming the BOP matrix $M_{BOP}$, the modified TF-DF method converts the frequency of each SAX word to a weight by counting the number of times a SAX word occurred in a PMU voltage time series and the number of PMU time series containing the SAX word. As a result, a weighted matrix $W_{BOP}$ is constructed. This matrix assigns a higher weight to a SAX word that represents most of the PMU voltage time series under each transient event. A column vector of voltage features can be extracted from the weighted matrix $W_{BOP}$, where the element of the voltage feature vector is the mean of the corresponding row of $W_{BOP}$, which is given as:

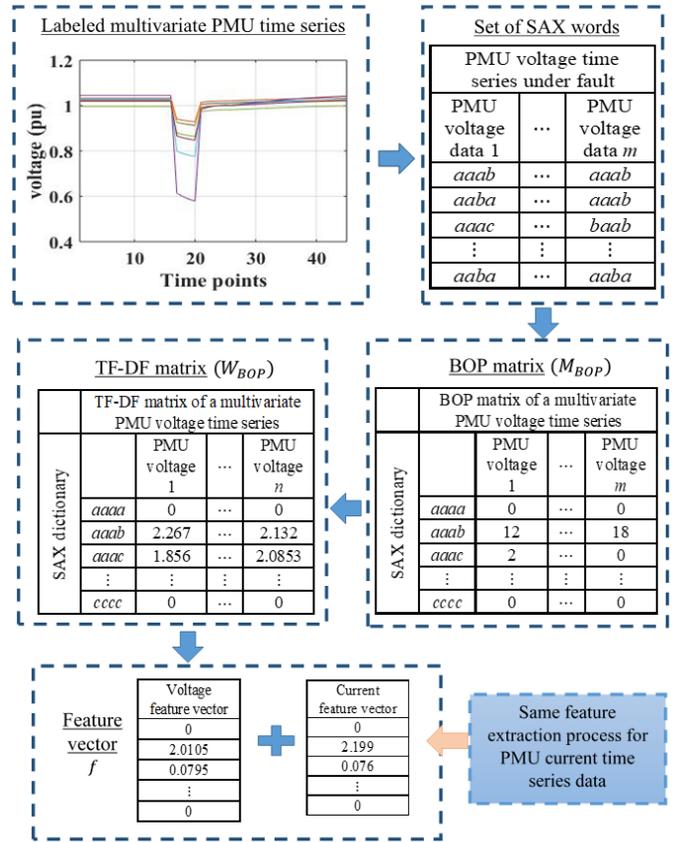

Fig. 5 Training phase of the multivariate BOP model

$$f_{voltage,i} = \frac{\sum_{j=1}^{n} W_{BOP_{i,j}}}{n} \quad (6)$$

where $f_{voltage,i}$ is the $i^{th}$ element of the voltage feature vector, and $n$ is the total number of columns in the weighted matrix $W_{BOP}$. The PMU current feature vector can be extracted from the labeled multivariate PMU current time series data following the same feature extraction process. Finally, the labeled multivariate PMU time series' feature vector, $f^T = [f_{current}^T, f_{voltage}^T]$, is constructed by combining the voltage feature vector $f_{voltage}$ and current feature vector $f_{current}$ together.

An example of the feature vectors' histogram distribution for each transient event is plotted in Fig. 6. It is seen that different events illustrate a different distribution for feature vectors. This is especially true for false data injection. Almost all the feature vector elements in a false data injection scenario range from 0 to 1, while the distribution of fault feature vector ranges from 0 to 2.5. This is due to the nature of the false data injection is to change several PMU measurements while the remaining measurements are kept constant.

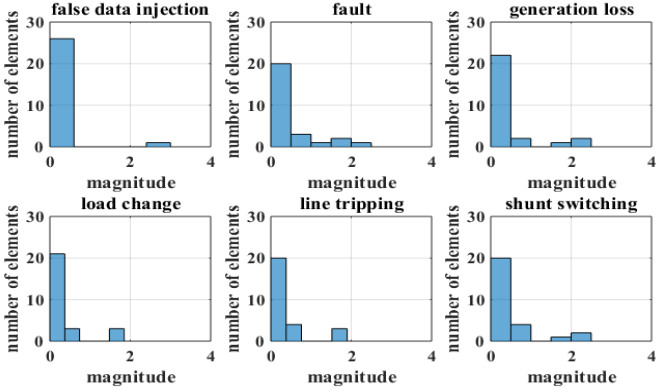

Fig. 6 Histogram of the feature vector for each event ($\alpha = 4$, $\gamma = 4$, $\omega = 25$)

### E. Classification and accuracy analysis

After extracting features from the labeled multivariate PMU time series data, SVM and Ensemble classifiers are used to recognize these six types of events. SVM enables transforming the data into a higher dimension so that an optimal hyperplane, which is the decision boundary, can be identified [28], [29]. High accuracy and the ability to handle complex nonlinear decision boundaries are the two main advantages of this classification method. In addition to SVM classifier, Ensemble method can achieve a high classification accuracy by using bag of trees as it applies multiple models to vote for the best result [28].

In our analysis, the aforementioned multivariate time series data containing 35 variables (28 current magnitudes and 7 voltage magnitudes) is used to classify faults, generation loss, load change, line tripping, shunt switching, and false data. As a result, the number of the columns of voltage BOP matrix and current BOP matrix are 7 and 28 respectively. The size of the BOP matrix's columns depends on the alphabet size $\alpha$ as well as the size of the word $\gamma$ in the SAX method.

The impact of SAX's parameters in classification accuracy is investigated by varying the size of alphabet, word, and the length of the sliding window. The classification accuracy, which distinguishes false data injection, fault, generation loss, line tripping, load change, and shunt switching, is evaluated after feeding the feature vectors into SVM and Ensemble classifiers with 10 fold cross validation, which is given in Fig.7. The classification results presented in Fig. 7 show that the multivariate BOP method performs well on classifying these six types of events. The proposed method achieves 99.3% accuracy on this 1.5s data when the alphabet size is 4, size of word is 4 and the length of sliding window is 25. The classification results from confusion matrix given in Fig. 8 show all false data, fault and shunt switching are correctly classified. One percent of load change is misclassified as generation loss while another three percent of load change is misclassified as line tripping. This is due to the similarity of the transient data between these three events. Nearly 99% line tripping and 99% generation loss are correctly classified. Further analysis of the results presented in Fig. 7 shows that increasing the length of sliding window and the size of the word will lead to an improvement in accuracy. However, there is a trade-off between the size of the feature vector and the size of the word. The longer the word size, the larger would be the feature vector. A large feature vector leads to a longer computation time for classification. This is not desired for real-time applications, where the time it takes to ident-ify events should be minimized. Comparing the accuracy and the size of the feature vector, a combination where alphabet size is four, size of the word is four, and the sliding window is 25 is proposed for classifying events as the classification accuracy reaches 99.3%. It is observed that SVM classifier outperforms Ensemble classifier by comparing the accuracies in Fig. 7.

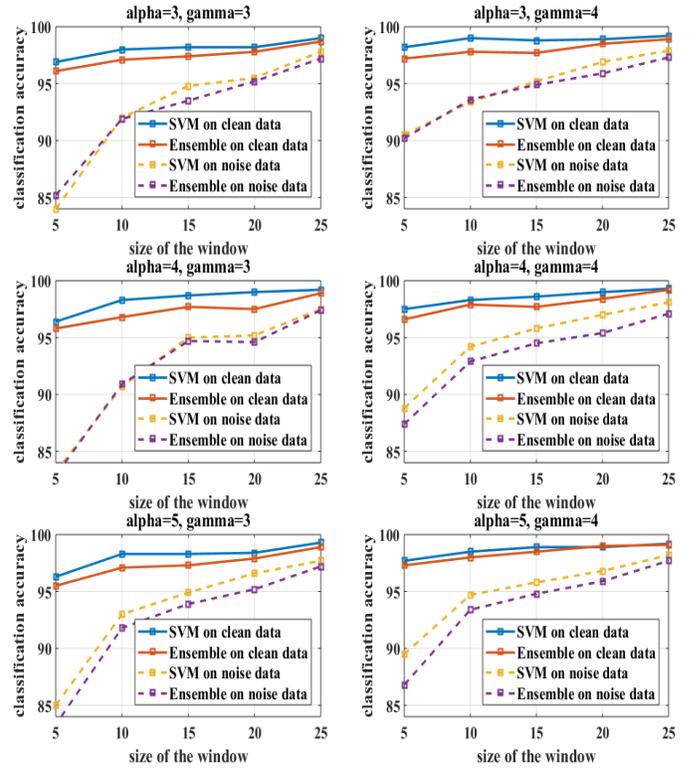

Fig. 7 Average classification accuracy (in percent) of six events for different SAX parameters

| | False Data | Fault | Generation Loss | Load Change | Line Tripping | Shunt Switching |
|---|---|---|---|---|---|---|
| False Data | 100% | | | | | |
| Fault | | 100% | | | | |
| Generation Loss | | | 99% | 1% | | |
| Load Change | | | | 96% | 1% | |
| Line Tripping | | | 1% | 3% | 99% | |
| Shunt Switching | | | | | | 100% |
| Positive Predictive Value | 100% | 100% | 99% | 96% | 99% | 100% |

Fig. 8 Confusion matrix for multivariate BOP model (accuracy = 99.3%, $\alpha = 4$, $\gamma = 4$, $\omega = 25$)

Variations in number of PMU as well as the location of PMUs are also considered in this paper to evaluate the effectiveness of the developed methodology. Results presented in Table V and Table VI show that the multivariate BOP model achieves a high accuracy with different coverage rates and placements of PMUs.

The amount of noise present in the PMU field data decreases the accuracy of classifying transient events. Thereby, it is important to either eliminate the noise or guarantee robustness of classification methods to noise. As indicated in [8], it is hard to filter out noise due to similarity between noise and power system fluctuations such as load change. This emphasizes the necessity to investigate robustness of the proposed BOP method aga-inst noise. Xie et al. in [31] propose adding a noise with 92 dB signal to noise ratio (SNR) to simulated PMU data so that the combined data resembles a field PMU data. As a result, a white Gaussian noise with 90 dB SNR is injected to all 2,353 studied event data. The proposed method achieves a 98.1% accuracy of classification with the previously suggested SAX parameters. Intuitively, the mean value computed from each subsequence reduces the influence of the white Gaussian noise.

TABLE V
CLASSIFICATION ACCURACY UNDER DIFFERENT COVERAGE OF PMUS ($\alpha = 4$, $\gamma = 4$, $\omega = 25$)

| Strategy | PMU placement | Classification accuracy (%) |
|---|---|---|
| 5PMUs | 2 6 10 12 27 | 99.5 |
| 6PMUs | 2 6 10 12 15 27 | 99.2 |
| 7PMUs | 2 4 6 10 12 15 27 | 99.3 |

TABLE VI
CLASSIFICATION ACCURACY UNDER DIFFERENT PLACEMENT OF PMUS ($\alpha = 4$, $\gamma = 4$, $\omega = 25$)

| Strategy | PMU placement | Classification accuracy (%) |
|---|---|---|
| 5PMUs | 2 6 10 12 27 | 99.5 |
| 5PMUs | 4 6 7 15 25 | 98.9 |
| 5PMUs | 2 4 15 17 28 | 98.7 |

TABLE VII
ACCURACY OF VARIOUS METHODS FOR BOTH CLEAN AND NOISY DATA

|  | Accuracy (%) Clean data | | Accuracy (%) Noisy data | |
|---|---|---|---|---|
|  | SVM | Ensemble | SVM | Ensemble |
| PCA | 51.3 | 52.3 | 29.5 | 29.4 |
| DWT | 87.5 | 93.2 | 76.6 | 91.3 |
| Dshapelet | 61.6 | 92 | 54.7 | 87.5 |
| $S^3$ | 65.1 | 96.4 | 77.9 | 89.9 |
| BOP | 99.3 | 99.2 | 98.1 | 97.1 |

Classification accuracies of traditional pattern recognition methods such as Discrete Wavelet Transform (DWT), Principal Component Analysis (PCA), Domain Specific Shapelet (Dshapelet) and Slope of Dshapelet ($S^3$) are compared with the developed multivariate BOP method. DWT uses the mother wavelet, which is a set of basic functions decomposing the data into multiple components, to transform the time series data into several resolution levels [32]. The coefficients, which represent the detailed and approximate information of the data, can be used as features for pattern recognition [33]. The mother wavelet Daubechies and all the approximate and detailed coefficients are selected to classify transient events. PCA projects the data onto the principal subspace, which is a lower dimensional linear space, such that the variance of the data is maximized [33]. The dynamics of the data can be analyzed by transferring the data into a combination of Principal Components (PCs) [1]. Time series can be expressed to a set of PCs where the first PC represents the largest variance of the data. We use the largest 20 PCs as features to distinguish each event. Shapelet is the subsequence of a time series that can maximally represent a time series [10]. Authors in [1] and [8] utilize Dshapelet and $S^3$ to extract the shapelet of the time series data and the one-step slope sequence of the shapelet as the feature vector. Dshapelet identifies the unique shapelet that can represent the most dramatic change while $S^3$ captures the trend of the most dramatic change [1], [8]. SVM and Ensemble classifiers are used to compare these traditional methods on clean data and simulated field data. Classification accuracies are given in Table VII and indicate that the developed multivariate BOP method leads to a better accuracy when classifying the aforementioned six types of events and is more robust against noise. Feature extraction time for PCA, DWT, Dshapelet, $S^3$ and multivariate BOP method are 0.009s, 0.005s, 0.0015s, 0.002s and 0.026s respectively, on a computer with i7-7700 CPU, 4.2GHz core and 32GB RAM. Compared to the running cycle of state estimation, which is approximately 30s, the feature extraction time can be neglected.

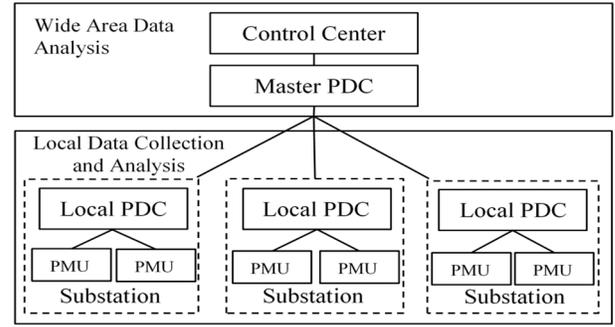

Fig.9 Distributed wide area monitoring systems

*F. Potential Application for larger power systems*

The proposed multivariate BOP model is shown to be effective with PMUs in IEEE 30-bus system. However, in a real-world scenario, where hundreds of PMUs are deployed, constructing a BOP matrix can prove to be cumbersome. However, this challenge can be addressed through deploying Wide Area Monitoring Systems (WAMS). WAMS are state monitoring systems that employ real time data obtained from PMUs to evaluate power system states and assess system transient stability [34]. Utilizing WAMS, the developed approach can be applied to each local PDC, since the total number of incoming PMU measurements to a PDC is tractable. In addition, post-disturbance analysis can be performed by WAMS. Distributed WAMS can be divided into two main parts as plotted in Fig. 9: wide area data analysis, and local data collection and analysis [35]. The PMUs located at substations send GPS time stamped phasor data to their local PDC where phasor data will be processed to stream. The stream out of local PDC will be sent to a master PDC for system monitoring and stability analysis.

For faster transient event detection, the time for event classification should be as short as possible. However, in large power system where hundreds of PMUs are deployed, constructing only two BOP current and voltage matrixes that consist of thousands of voltage and current PMU measurements is time consuming. Since the proposed BOP approach is independent of the coverage rate and location of PMUs, it is possible to add another layer to local PDC installed in each substation. Instead of constructing only two large BOP matrixes, each local PDC can perform the multivariate BOP model individually and can use the collected substation's voltage and current measurements to determine the correctness of the substation's PMU data and identify the type of transient events that have occurred in the system. When an event is detected, the information can be transformed to master PDC to inform the control center of the power grid operating conditions within milliseconds. Intuitively, each local PDC shares the task of event detection so that the applicability of the multivariate BOP method is ensured in larger power systems.

## VI. Conclusions

In this paper, a structural-based multivariate BOP model is proposed to use PMU time series data to detect false data and classify transient events. An efficient data reduction technique, i.e. SAX, is utilized to prune and convert PMU time series into words such that the event patterns can be recorded in a multivariate BOP frequency matrix and features can be extracted by applying the modified TF-DF method. Results show that the multivariate BOP model is efficient on this short 1.5s time series data and outperforms other classical methods in detecting false data and classifying transient events in terms of classification accuracy and robustness against noise. The independence of system topology and loading conditions as well as the coverage rates and locations of PMUs guarantees the BOP method's applicability in large power system. This BOP method provides a fast event detection tool that can be used to inform the control center of the power grid operating conditions and thus improve the power system situational awareness and reliability.